\documentclass[11pt,twoside]{article}

\usepackage{asp2006}
\usepackage{epsf}
\usepackage{psfig}
\usepackage{lscape}

\newcommand{\kms}{km s$^{-1}$}

\newcommand{\mass}{$M_\odot$}

\newcommand{\ha}{H$\alpha$}

\begin{document}
\title{Dwarf Detachment and Globular Cluster Formation in Arp 305}   
\author{Hancock, M.$^1$, Smith, B. J.$^2$, Struck, C.$^3$, Giroux, M. L.$^2$, Hurlock, S.$^2$}   
\affil{$^1$University of California, Riverside; $^2$East Tennessee State University; $^3$Iowa State University}    

\begin{abstract} 
Tidal Dwarf Galaxies (TDG), concentrations of interstellar gas and
stars in the tidal features of interacting galaxies, have been the
subject of much scrutiny.   The `smoking gun' that will prove the TDG
hypothesis is the discovery of independent dwarf galaxies that are
detached from other galaxies, but have clear tidal histories.  As part
of a search for TDGs we are using GALEX to conduct a large UV imaging
survey of interacting galaxies selected from the Arp Atlas.   As part
of that study, we present a GALEX UV and SDSS and SARA optical study
of the gas-rich interacting galaxy pair Arp 305.  The
GALEX UV data reveal much extended diffuse UV emission and star
formation outside the disks including a candidate TDG between the two
galaxies.  We have used a smooth particle hydrodynamics code to model
the interaction and determine the fate of the candidate TDG.
\end{abstract}



\section{INTRODUCTION}

The so-called `Tidal Dwarf Galaxies' (TDG), concentrations of
interstellar gas and stars in the tidal features of interacting
galaxies that may become independent dwarf galaxies, have been the
subject of intense scrutiny (e.g., \citealp{bou06,rec06,duc06}).
The `smoking gun' that will unambiguously prove the TDG
hypothesis is the discovery of independent dwarf galaxies that are
detached from other galaxies, but have clear tidal histories.

As part of a search for TDGs and to study star formation in tidal
features, we have used the {\it Galaxy Evolution Explorer} (GALEX) telescope
\citep{mar05} to conduct a large UV imaging survey of interacting
galaxies selected from the Arp\ (1996) Atlas. 
We have found a number of  previously
unstudied candidate TDGs in our sample (\citealp{smi09}, 
also see B. Smith et al., this conference proceedings).

In this proceedings we present results from \citet{han09} 
on the interacting galaxy pair Arp 305 (NGC
4016/7).  We have obtained UV and optical images of Arp 305 from the
GALEX, Sloan Digitized Sky Survey (SDSS), and the Southeastern
Association for Research in Astronomy (SARA) telescopes.  

\section{DISCUSSION}

\subsection{Morphology and General Properties}

Arp 305 is a very wide pair with the primary galaxy, NGC 4017, to the
South and the companion, NGC 4016, to the North.   
NGC 4017 appears nearly face-on, with two tidal
tails, one pointing northwest and one to the southeast.  The UV images
show much extended emission to the northwest and southeast far outside
the main disk \citep{han09}.   The primary seems to have an ocular waveform, a
bright oval of star formation shaped like an eyelid (e.g. 
\citealp{kau97,han07,elm06})  

The Northern galaxy, NGC 4016, shows a misshapen bulge with a dusty disk.  
In the inner disk, a curious figure-eight shape is seen \citep{han09}.  
We suspect that the
figure-eight formation is the result of a bar (see the
third frame of Figure 3 in \citealp{rom08}).  

\citet{elm93} noted that this galaxy pair showed scattered debris
resembling dwarf galaxies.  The most prominent debris is seen in a partial
residual bridge between the two spiral galaxies.  This feature is
particularly striking in the GALEX images (Figure 1).  For simplicity,
we will adopt the name `bridge TDG' for the tidal dwarf candidate in
the residual bridge.  With SARA, we have detected \ha\ emission from
this feature confirming that it is at the same redshift as the
galaxies.   This structure is clearly detected in HI
\citep{moo83}, further suggesting that it is part of the Arp 305
system.

\subsection{Reddening and Ages}

We have identified 45 star forming clumps in Arp 305 from the GALEX
FUV images.  The bridge TDG contains four of these clumps (Figure 1).
Two additional detached TDG candidates northeast and
southwest of NGC 4016 are also marked in Figure 1.

To estimate the ages of the star forming clumps we 
compared various color combinations to sets of Starburst99 (SB99)
models \citep{lei99}.  See \citet{han09} for details on the SB99 models
and the age determinations.

We do not see any old clumps in this system in spite of the fact that
the last closest encounter was about 300 Myr ago.   Perhaps this is
the  result of cluster dissolution or the so called cluster `infant
mortality' (e.g. \citealp{pet09,bas05,fal05}).  However, this could be
a selection  effect.  We selected the clumps from the FUV image so
have chosen the youngest clumps.  Another possibility is our limited
resolution.  It is likely that the clumps are made up of several
unresolved clusters and/or associations.  The light in our photometric
apertures would be dominated by the younger  clusters.

The absence  of intermediate
age clumps in the tidal structures of Arp 305 and some other systems
(e.g, Arp 82 and Arp 285; \citealp{han07,smi08}) indicates that it is
difficult to make long-lived TDGs.

\subsection{The Bridge TDG}

The bridge TDG looks to be embedded in a massive HI plume stretching
North  from the primary \citep{moo83}.  
The stars in the bridge TDG most likely formed in situ.
The distance between
the nucleus of the primary and the bridge TDG is $\sim36$ kpc.  
The clumps in the bridge TDG are an average of 19 Myr
old.  To travel 36 kpc in 19 Myr the material would have to travel at
a velocity of roughly 2000 \kms\ relative to the primary.  The HI
kinematics do not support this \citep{moo83}.

We estimate that the combined stellar mass for the clumps in the
bridge TDG is $\sim1-7\times10^{6}$\mass.  
These clump masses
are consistent with that of Galactic globular clusters (e.g.,
\citealp{pry93}) and  Super Star Clusters (SSC)
(e.g. \citealp{hol92,hol96,oco94,sch96,whi93,whi95,wat96}).   For
comparison, the TDG candidates in \citet{hig06} have stellar masses around
2$\times10^7$\mass\ to 3$\times10^8$\mass.

The HI mass of the bridge TDG is $\sim6\times10^{7}$\mass\ 
consistent with the 29 low mass dwarfs studied in \citet{beg08} ($10.18 -
81.14\times10^6$\mass).  
The HI mass/luminosity ratio for the bridge TDG is M$_{HI}$/L(B)$\sim1$
M$_{\odot}$/L$_{\odot}$, similar to those of irregular and compact
blue dwarfs (e.g. \citealp{hun04,pis05,tar05,beg08}).  The bridge TDG
has an \ha\ luminosity of $3.2\times10^{39}$ erg s$^{-1}$, similar
to the \ha\ luminosities of the dwarfs in \citet{hun04} and the tidal
features studied by \citet{smi01}.

\subsection{Other Clumps of Interest}

There are two other TDG candidates in Arp 305.
To the southwest of NGC 4016 is a bright clump (Figure 1), that appears
to be at the tip of a faint tidal arm.   This clump is bright in both
the GALEX UV and all the SDSS optical bands and is within the extended
HI envelope \citep{moo83}.   It is not detected in  our continuum
subtracted \ha\ images. This TDG candidate is the oldest clump in our
sample.  We can not rule out the possibility that this
object is a foreground star or a background object.

To the northeast is another possible TDG.  The northern tail
of NGC 4016 points toward this clump.  This TDG candidate is also  bright in
both the GALEX UV and all the SDSS optical bands, and is not detected
in our continuum subtracted \ha\ images.
Figure 1 shows this northern TDG candidate to be an extended object
with an appearance similar to an inclined disk.  Given this, the lack
of an \ha\ detection, and the large distance from NGC 4016 ($\sim70$
kpc), we can not rule out the possibility that this is a
background galaxy.

Near the bases of both the northeastern and southwestern tidal tails
in NGC 4017 are extremely luminous clumps.  These
`hinge clumps' likely form when material is pulled out from  deeper in
the original disk.  This material, gas with  higher initial densities,
is more compressed.  Moreover, being pulled out  in a tail likely
reduces the shear levels of the original disk, allowing self-gravity
to more easily form big clouds.  The `hinge clumps' are the two
brightest UV clumps in the primary and are very bright in \ha.  
A luminous `hinge clump' was also observed  in Arp 82 at the base of the long
extended northern tail \citep{han07}.

\subsection{SPH Model of the Encounter}

What is the long term fate of the clusters that do survive, and do any
of those that might represent TDGs detach in some sense?  We have
investigated those questions by running  numerical simulations.  We
used the SPH code of \citet{str97}.  This code was also used to model
other systems recently (e.g. Arp 284, \citealp{str03}; IC 2163/NGC
2207,  \citealp{str05}; Arp 107, \citealp{smi07}; Arp 82,
\citealp{han08};  and Arp 285, \citealp{smi08}).

The TDG particle in Figure 2 turned on SF at a
time and place like that of the bridge TDG in the observations, and might
illustrate the fate of the bridge TDG.  If so, that fate is to be
captured by the companion and carried into the merger.  It seems very
unlikely that TDGs formed in the bridge will detach and survive the
merger.

The models suggest the eventual merger is about 1.3 Gyr from the time 
depicted in Figure 2.  
The bridge TDG could persist for this long, consistent with recent results.
In their extensive N-body study of tidal dwarf formation \citet{bou06}
found that 75\% of the dwarf candidates fell back into the
galaxies within a few $\times10^8$ yr.  The remaining 25\% had a typical
lifetime of more than 2 Gyr.  Most of the later formed in the outer
parts of tidal tails.
However, we are not sure if the bridge TDG is currently bound.  We don't 
know whether it will survive the first and subsequent fall backs.

\section{SUMMARY}

We present results from \citet{han09} on the candidate TDGs in the 
interacting galaxy pair Arp 305 (NGC 4016/7).
A prominent TDG candidate is seen in a partial residual bridge between the 
two spiral galaxies.  We summarize the bridge TDG analysis below:

\begin{itemize}
\item Mean Age$\sim$19 Myr, E(B-V)$\sim$0.06
\item Stars formed in situ
\item Total Stellar Mass of the 4 clumps $\sim1-7\times10^{6}$\mass 
\item Total HI Mass $\sim6\times10^{7}$ \mass
\item Mass more similar to SSCs or perhaps a small TDG
\item Has a clear tidal history
\item It is not clear if the TDG is a single bound object
\item Unlikely to permanently detach and survive the merger
\item Could persist for $\sim1.3$ Gyr
\end{itemize}

\acknowledgements 
This research was supported by
NASA  LTSA grant NAG5-13079 and GALEX grant GALEXGI04-0000-0026.
This work has made use of the NASA/IPAC Extragalactic Database (NED),
which is operated by the Jet Propulsion Laboratory, California
Institute of Technology, under contract with NASA.


\begin{figure*}[ht]
\begin{center}
\scalebox{0.4}{\rotatebox{0}{\includegraphics{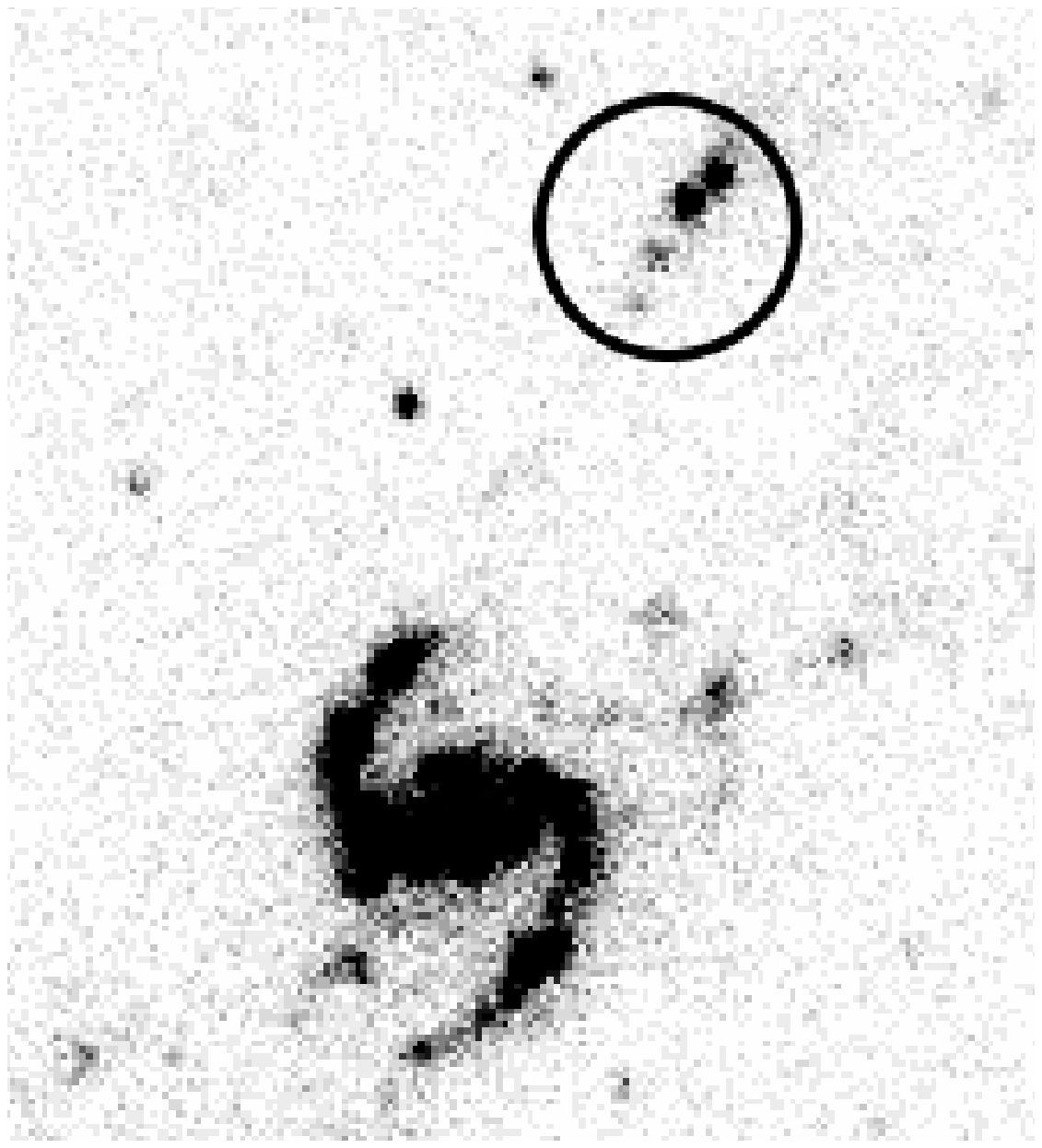}}}
\scalebox{0.3}{\rotatebox{0}{\includegraphics{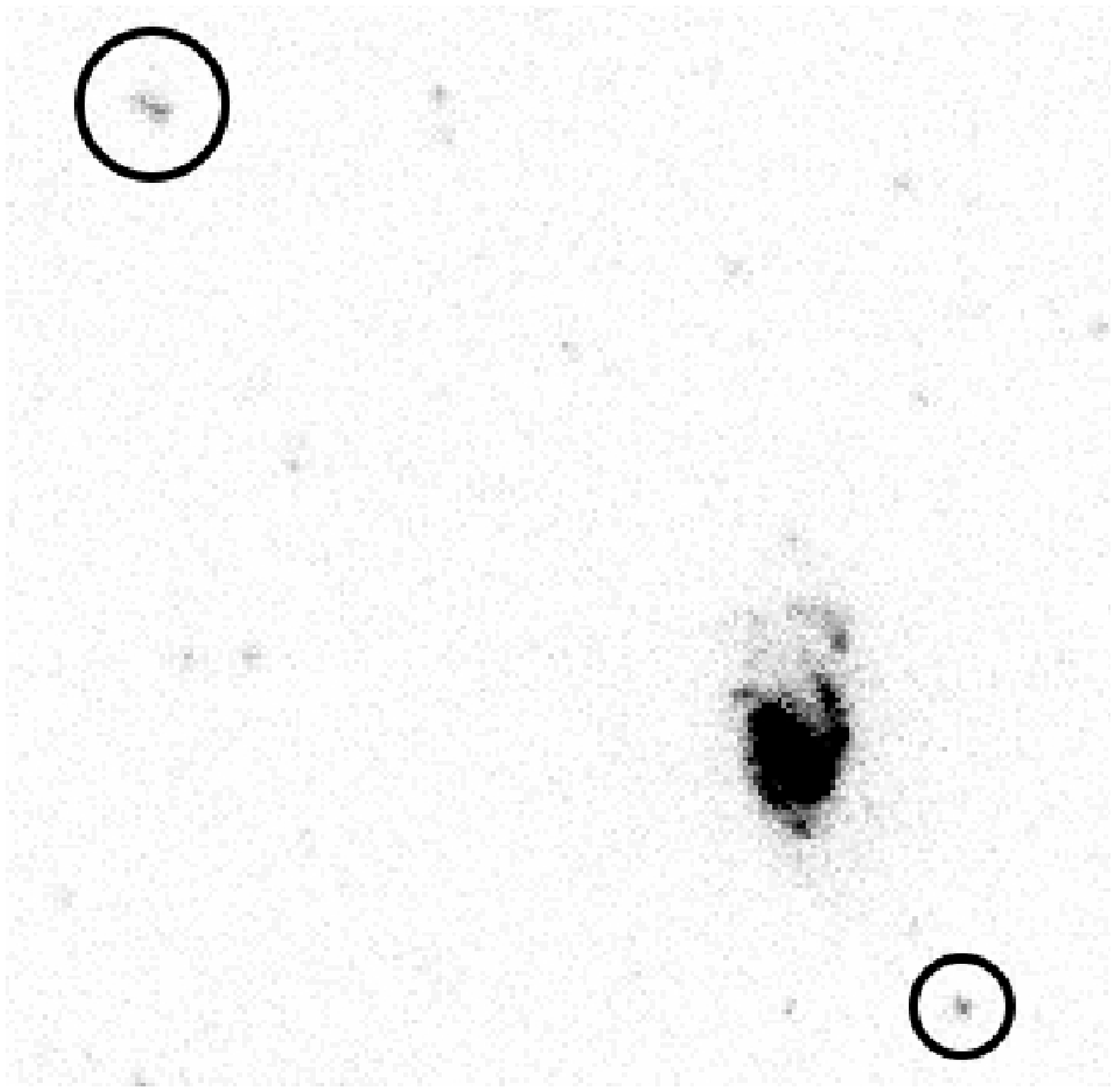}}}
\end{center}
\caption[]{GALEX FUV images of Arp 305.  Left:  NGC 4017 with
the bridge TDG circled.  Right:  NGC 4016 with two candidate TDGs
circled.  North is up and East is to the left.  See \citet{han09}
for color images with scale bars.}
\end{figure*}

\begin{figure*}[ht]
\begin{center}
\scalebox{0.6}{\rotatebox{0}{\includegraphics{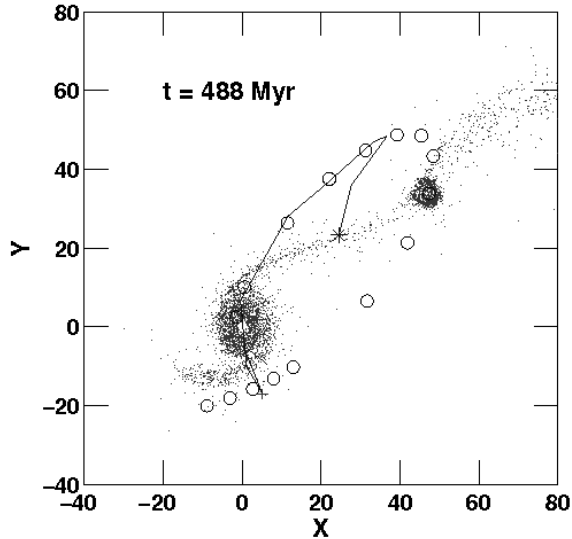}}}
\end{center}
\caption[]{Illustration of the orbital trajectory of star-forming
  particles from the onset of star formation in the bridge to the end
  of the run when the galaxies have merged, over-plotted on the gas
  particle distributions at a selected time in the interaction.  Dots
  show the locations of gas particles, at the time of SF onset in the
  selected particles.  Black circles show the position of the
  companion center at selected time-steps from closest passage to
  merger.  The asterisk shows star-forming particles in selected tidal
  structures, i.e. the TDG.   The curve shows the particle trajectory
  from the onset of SF to the end of the run when the two galaxies
  have merged.  }
\end{figure*}

\end{document}